\documentstyle[preprint,prd,aps]{revtex}
\begin{document}
\draft

\title{$\bf U(1,1)$--Invariant Generation of Charges for
Einstein--Maxwell--Dilaton--Axion Theory}

\author{Oleg Kechkin and Maria Yurova}

\address{Nuclear Physics Institute,\\
Moscow State University, \\
Moscow 119899, RUSSIA, \\
e-mail: kechkin@cdfe.npi.msu.su}

\date{August 1996}

\maketitle

\draft

\begin{abstract}
The action of the isometry subgroup which preserves the trivial values of the
fields is studied for the stationary D=4 Einstein--Maxwell--Dilaton--Axion
theory. The technique for generation of charges and the corresponding
procedure for construction of new solutions is formulated. A solution
describing the double rotating dyon with independent values of all physical
charges is presented.
\end{abstract}



\draft

\narrowtext

\section{Introduction}
The low--energy heterotic string theory describes, apart from gravitational,
dilaton and axion fields, some additional scalar and vector ones, originated
from the compactification of extra dimensions. The stationary variant
of this theory admits the chiral matrix representation \cite {ms}--\cite {mahs},
and becomes completely integrable after imposing of the axisymmetric
condition \cite {bkr}--\cite {mah}.

The Einstein--Maxwell--Dilaton--Axion (EMDA) theory, being the simplest
system of this type, allows an invariance under the $Sp(4,R)$ group of
isometry transformations for its $Sp(4,R)/U(2)$ target space in the
stationary case \cite {ggk1}--\cite {gk1}. It has been established a
remarkable formal analogy between the stationary EMDA and pure Einstein
theories, which can be directly prolonged to the stationary and axisymmetric
case \cite {ky1}.

In this paper we study the maximal isometry subgroup preserving the trivial
values of the fields for the stationary EMDA theory. We present the
subgroup action on the EMDA Ernst--like potentials, combined into $2\times 2$
symmetric complex matrix. This action becomes especially simple for the
Coulomb expansion near the space infinity or, equivalently, for the physical
charges of the model. Using the charge matrix, corresponding to the potential
one, we show that it undergoes linear transformations with a belonging
to $U(1,1)$ matrix operator.

It is established, that this operator preserves two quadratic forms,
constructed on the charges: the first one which is connected with the
Bogomol'nyi bound, and the second one, which is related to the well known
charge restriction for EMDA black hole solutions. Then, the charge generation
technique allowing to obtain charges compatible with the Bogomol'nyi bound,
as well as the corresponding solution construction procedure, is presented.

In the last section of the paper the above formulated approach is applied
to the construction of the solution, possessing independent values of all EMDA
charges and describing the double rotating dyon. This solution contains the
one obtained in \cite {gk2}, which has trivial values of NUT, magnetic and
axion charges, and also generalizes the single dyon solutions found before
\cite {gk3}--\cite {ggk2}.
\section{Target Space Transformations}
The EMDA theory in four dimensions has the action
\begin{eqnarray}
S = \int d^4x {\mid g \mid}^{\frac {1}{2}} (-R+2{\partial \phi}^2+
\frac {1}{2}e^{4\phi }{\partial \kappa}^2  -e^{-2\phi}F^2 -
\kappa F\tilde {F}),
\end{eqnarray}
where $\phi$ is the dilaton field, $\kappa$ is the axion and for the
electromagnetic tensor $F_{\mu \nu}$ we put
\begin{eqnarray}
F_{\mu \nu} = \partial _{\mu}A_{\nu}-\partial _{\nu}A_{\mu},
\qquad
\tilde {F}^{\mu \nu} = \frac {1}{2} E^{\mu \nu \lambda \sigma}
F_{\lambda \sigma}.
\end{eqnarray}

It is convenient to parametrize the four--dimensional line element
as \cite {iw}
\begin{equation}
ds^2=f(dt-\omega _idx^i)^2-f^{-1}h_{ij}dx^idx^j,
\end{equation}
where $i=1,2,3$.
Below we will deal with the stationary case, when it is possible to
introduce the magnetic $u$ and rotational $\chi$ potentials \cite {ggk1}:
\begin{eqnarray}
\nabla u &=& fe^{-2\phi}(\sqrt{2}\nabla \times \vec A
+\nabla v \times \vec \omega)+\kappa \nabla v,
\nonumber\\
\nabla \chi &=& u\nabla v-v\nabla u -f^2\nabla \times \vec \omega
\end{eqnarray}
(here $v=\sqrt{2}A_0$, and operator $\nabla$, as any other 3--vector
variable, is defined using the metric $h_{ij}$). Then the complete set of
equations can be derived from the action \cite {gk1}
\begin{equation}
^3S=\int d^3x h^{\frac {1}{2}}(-^3R+\frac {1}{4}TrJ_M^2),
\end{equation}
where $^3R$ is the  curvature scalar constructed using $h_{ij}$,
and $J_M=\nabla M M^{-1}$. Here the $4\times 4$ matrix $M$ is defined by
two $2\times 2$ symmetric matrices $P$ and $Q$,
\begin{eqnarray}
M=\left (\begin{array}{crc}
P^{-1}&P^{-1}Q\\
QP^{-1}&P+QP^{-1}Q\\
\end{array}\right ),
\end{eqnarray}
which explicit form is:
\begin{eqnarray}
P = \left (\begin{array}{crc}
f-v^2e^{-2\phi}&-ve^{-2\phi}\\
-ve^{-2\phi}&-e^{-2\phi}\\
\end{array}\right ),\quad
Q = \left (\begin{array}{crc}
-\chi +vw&w\\
w&-\kappa\\
\end{array}\right ),
\end{eqnarray}
where $w=u-\kappa v$ \cite {gk2}.

A symmetric matrix $M$, being also a symplectic one,
\begin{eqnarray}
M^TJM=J,\quad
{\rm where}\quad
J=\left (\begin{array}{crc}
0&-I\\
I&0\\
\end{array}\right ),
\end{eqnarray}
belongs to $Sp(4,R)/U(2)$, and remains in it under the map
\begin{equation}
M\rightarrow G^TMG
\end{equation}
for an arbitrary symplectic matrix $G$.
Such transformations form the isometry
group $Sp(4,R)$ for the target space of the stationary EMDA theory.

Here we study the action of the isometry subgroup, which
preserves the trivial values of the fields. The relation defining this
subgroup, according to (6), (7) and (9), is
\begin{eqnarray}
\Sigma _3=G^T\Sigma _3G,\quad {\rm where}\quad
\Sigma _3=\left (\begin{array}{crc}
\sigma _3&0\\
0&\sigma _3\\
\end{array}\right )
\end{eqnarray}
is the trivial value of the matrix $M$, and $\sigma _i$ are the Pauli
matrices.

It can easily be shown, that the relations (8) and (10) allow to obtain
the explicit form of the matrix $G$. The result is:
\begin{equation}
G=G_{(0)}G_{(3)}=G_{(3)}G_{(0)},
\end{equation}
where
\begin{eqnarray}
G_{(0)}=I\cos \lambda ^0+\Gamma _0\sin \lambda ^0,
\qquad
G_{(3)}=f_1+\lambda ^i\Gamma _if_2,
\end{eqnarray}
with
\begin{eqnarray}
2f_1 &=& (1+\sigma)\cosh \alpha +(1-\sigma)\cos \alpha,
\nonumber \\
2\alpha f_2 &=& (1+\sigma)\sinh \alpha +(1-\sigma)\sin \alpha.
\end{eqnarray}

Here $\lambda ^{\mu}$ are four real parameters,
$\alpha^2=\sigma \eta _{ij}\lambda ^i \lambda ^j$,
$\sigma =sign(\eta _{ij}\lambda ^i \lambda ^j)$ and
$\eta _{ij}=diag(1,1,-1)$.

The four matrices $\Gamma _{\mu}$,
\begin{eqnarray}
\Gamma _0=\left (\begin{array}{crc}
0&\sigma _3\\
-\sigma _3&0\\
\end{array}\right ),\quad
\Gamma _1=\left (\begin{array}{crc}
-\sigma _1&0\\
0&\sigma _1\\
\end{array}\right ),\quad
\Gamma _2=\left (\begin{array}{crc}
0&\sigma _1\\
\sigma _1&0\\
\end{array}\right ),\quad
\Gamma _3=\left (\begin{array}{crc}
0&\sigma _0\\
-\sigma _0&0\\
\end{array}\right ),\quad
\end{eqnarray}
where $\sigma _0=I_2$,
are the generators of the gl(2,R) algebra, and the general subgroup matrix
(11) realizes the four--dimensional GL(2,R) representation.

It can be easily proved, that a matrix allowing the Gauss decomposition
\begin{eqnarray}
G=\left (\begin{array}{crc}
{S^T}^{-1}&{S^T}^{-1}R\\
L{S^T}^{-1}&S+L{S^T}^{-1}R\\
\end{array}\right ),
\end{eqnarray}
where $L$ and $R$ are symmetric $2\times 2$ matrices, possesses
the symplectic property (but, for example, the symplectic matrix $J$ can
not be represented in this form). The fact is that both matrices $G_{(0)}$
and $G_{(3)}$ admit such decomposition. Namely, using direct calculation
one obtains, that
\begin{equation}
S_{(0)}=(\cos \lambda ^0)^{-1}\sigma _0,\qquad
R_{(0)}=-L_{(0)}=\sigma _3\tan \lambda ^0;
\end{equation}
and
\begin{eqnarray}
S_{(3)} &=& \Delta ^{-1}[f_1\sigma _0+\lambda ^1f_2\sigma _1],
\nonumber \\
R_{(3)} &=& f_2\Delta ^{-1}[(\lambda ^3f_1+\lambda ^1\lambda ^2f_2)\sigma _0+
(\lambda ^2f_1+\lambda ^1\lambda ^3f_2)\sigma _1],
\nonumber \\
L_{(3)} &=& f_2\Delta ^{-1}[(-\lambda ^3f_1+\lambda ^1\lambda ^2f_2)\sigma _0+
(\lambda ^2f_1-\lambda ^1\lambda ^3f_2)\sigma _1],
\end{eqnarray}
where $\Delta =f_1^2-(\lambda ^1)^2f_2^2$.

Let us now unit two real matrices $P$ and $Q$ into a complex one:
\begin{equation}
Z=Q+iP.
\end{equation}
Then the transformation (9) can be rewritten in the form
\begin{equation}
Z=S^T({Z_0}^{-1}+L)^{-1}S+R,
\end{equation}
where $Z_0$ denotes the initial value of the matrix $Z$. Thus, to obtain the
action of the $GL(2,R)$ group on the non--chiral matrix $Z$, one needs to
use (19) twice with $S$, $R$ and $L$ giving by (16) and (17). (Here a
concrete order is not important in view of (11)).

It is useful to establish the explicit transformation form in three special
cases. The first one is related with
$\{\lambda ^{\mu}\}=\{\lambda ^0, 0, 0, 0\}$, when
\begin{equation}
Z=H_{(0)}^{-1}[(1+(\tan \lambda ^0)^2)Z_0+\tan \lambda ^0
(1-W_0-(E_0+z_0)\tan \lambda ^0)\sigma _3],
\end{equation}
where $H_{(0)}=1-\tan \lambda ^0(E_0+z_0)+(\tan \lambda ^0)^2W_0$,
and $W_0=E_0z_0+\Phi _0^2$.
The second transformation is defined by
$\{\lambda ^{\mu}\}=\{0, 0, 0, \lambda ^3\}$:
\begin{equation}
Z=H_{(3)}^{-1}[(1+(\tan \lambda ^3)^2)Z_0+\tan \lambda ^3
(1+W_0-(E_0-z_0)\tan \lambda ^3)\sigma _0],
\end{equation}
where $H_{(3)}=1-\tan \lambda ^3(E_0-z_0)-(\tan \lambda ^3)^2W_0$.
Finally, the third special case is connected with
$\{\lambda ^{\mu}\}=\{0, \alpha \cos \beta, \alpha \sin \beta, 0\}$,
when
\begin{eqnarray}
Z &=& H^{-1}[Z_0+\sinh \alpha (2\sinh \alpha \Phi _0
+\cosh \alpha \Sigma _0)\sigma _1
\nonumber\\
&+& \cos \beta \sinh \alpha (2\cosh \alpha \Phi _0+\sinh \alpha \Sigma _0)
\sigma _0].
\end{eqnarray}
Here $H=1+\sin \beta \sinh \alpha (2\cosh \alpha \Phi _0+
\sinh \alpha \Sigma _0)$ and $\Sigma _0=(E_0-z_0)\cos \beta +
(1+W_0)\sin \beta$.

The isometry subgroup preserving the trivial values of the fields for
the low energy heterotic string theory with moduli fields was studied using
the real target space variables in \cite {cy1}.
\section{Charge Space Transformations}
Let us consider asymptotically trivial solutions of the EMDA equations. Then
near space infinity one has:
\begin{equation}
Z_{as}=i(\sigma _3-\frac {2\hat Q}{r}).
\end{equation}
Here $r$ denotes the asymptotical radial coordinate, and the matrix $\hat Q$
is constructed using three complex charges,
\begin{eqnarray}
\hat Q=\left (\begin{array}{crc}
{\cal M}&{\cal Q}\\
{\cal Q}&-{\cal D}\\
\end{array}\right ),
\end{eqnarray}
which are connected with the ADM mass, NUT parameter, and also with the
electric, magnetic, dilaton and axion charges:
\begin{eqnarray}
{\cal M} = M+iN,
\qquad
{\cal Q} = (Q_e+iQ_m)/\sqrt 2,
\qquad
{\cal D} = D+iA.
\end{eqnarray}

The employment of the belonging to GL(2,R) transformation does not change the
$Z_{as}$ form. The group relation (10), expressed using $R$, $L$ and $S$
matrices,
\begin{equation}
\sigma_3+iR=S^T(\sigma_3+iL)^{-1}S,
\end{equation}
ensures the equality $Z_{\infty}=i\sigma_3$. Then, straightforward calculation
leads to the following compact expression for a charge matrix transformation:
\begin{equation}
\hat Q=T^T\hat Q_0T.
\end{equation}
Here the matrix operator $T$ is defined as
\begin{equation}
T=(I+iL\sigma_3)^{-1}S,
\end{equation}
where $L$ and $S$ must be taken from (16) and (17).  The substitution
gives:
\begin{eqnarray}
T_{(0)} = e^{i\lambda^0}\sigma_0,
\end{eqnarray}
\begin{eqnarray}
T_{(3)} = f_1\sigma_0+f_2(\lambda^1\sigma_1-i\lambda^2\sigma_2+
i\lambda^3\sigma_3),
\end{eqnarray}
so that $[T_{(0)},T_{(3)}]=0$, and it is possible to define the general
operator $T$ as
\begin{equation}
T=T_{(0)}T_{(3)}=T_{(3)}T_{(0)}.
\end{equation}
This operator realizes the complex $2\times 2$ matrix representation for the
isometry subgroup $GL(2,R)$. The local isomorphism is defined
using the correspondence between the two sets of generators, $\{\Gamma _0,
\Gamma_1, \Gamma_2, \Gamma_3\}$ and $\{i\sigma_0, \sigma_1, -i\sigma_2,
i\sigma_3\}$.

One also can prove, that the following remarkable property takes place:
\begin{equation}
T^{+}\sigma_3T=\sigma_3,
\end{equation}
which means that the operator $T$ belongs to the $U(1,1)$ group. Moreover,
it is not difficult to show, that the operator defined by the formulae
(29)--(31) provides the complete realization of the $U(1,1)$ group.

It will be convenient to introduce the quadratic charge combination
\begin{eqnarray}
I_1 &=& Tr(\bar {\hat Q}\sigma_3\hat Q\sigma_3) =
\bar {\cal {M}}{\cal {M}}+\bar {\cal {D}}{\cal {D}}-2\bar {\cal {Q}}{\cal {Q}}
\nonumber\\
&=& M^2+N^2+D^2+A^2-Q_e^2-Q_m^2,
\end{eqnarray}
which also defines the Bogomol'nyi bound. Then, from (32) it immediately
follows, that $I_1$ is invariant under a $U(1,1)$ transformation.

The introduced norm $I_1$ is defined for an arbitrary charge space
point with coordinates $\{{\cal M}, {\cal D}, {\cal Q}\}$. Its complete
invariance group $U(2,1)$ acts transitively on the invariant surfaces of the
charge space defined by (33), but only the above described $U(1,1)$
subgroup is realized by isometry transformations preserving the trivial
values of the fields.

This subgroup, apart from the $I_1$ preserving, allows the additional
conservation property, as it follows from (32):
\begin{equation}
I_2 = \mid {\rm det} \hat Q \mid = \mid {\cal {M}}{\cal {D}}+{\cal {Q}}^2
\mid = {\rm inv}.
\end{equation}
Thus, if one starts from the initial vacuum solution with $z_0-i=\Phi_0=0$
and ${\cal {M}}_0\neq 0$, one obtains after the generation the well known
relation between the new physical charges \cite {gk3} and \cite {stw}:
\begin{equation}
{\cal {D}}=-\frac{{\cal {Q}}^2}{{\cal {M}}}.
\end{equation}
The same situation takes place for the case of $E_0-i=\Phi_0=0$ and
${\cal {D}}_0\neq 0$.

The way to get the ``fatal'' property (35) away is connected with the
original charge base extension and, correspondingly, with the generalization
of the initial solution class. Namely, it is easy to see, that written using
the $Z_0$--representation the stationary EMDA equations
\begin{eqnarray}
\nabla J_{Z_0} &=& J_{Z_0}(J_{Z_0}-\bar J_{Z_0}),
\nonumber\\
^3R_{ij} &=& 2Tr(J_{Z_0(i}\bar J_{Z_0j)}),
\end{eqnarray}
where $J_{Z_0}=\nabla Z_0(Z_0-\bar Z_0)^{-1}$, in the case of $\Phi _0=0$
reduce to the double Ernst system
\begin{eqnarray}
\nabla J_{E_0} = J_{E_0}(J_{E_0}-\bar J_{E_0}),
\qquad
\nabla J_{z_0} = J_{z_0}(J_{z_0}-\bar J_{z_0}),
\end{eqnarray}
\begin{eqnarray}
^3R_{ij} = 2Tr(J_{E_0(i}\bar J_{E_0j)}+J_{z_0(i}\bar J_{z_0j)}).
\end{eqnarray}
(Here $J_{E_0}$ and $J_{z_0}$ can be obtained from $J_{Z_0}$ by the
appropriate index replacement).
Then, in the stationary and axisymmetric case, after the
Lewis--Papapetrou three--dimensional line element parametrization,
\begin{equation}
dl^2 = h_{ij}dx^idx^j = e^{2\gamma }(d\rho^2+dz^2)+\rho^2d\varphi^2,
\end{equation}
both equations (37) become the flat two--dimensional Ernst one. Also
from (38) it follows, that $\gamma = \gamma _1+\gamma _2$, where the
functions $\gamma_1$ and $\gamma_2$ are defined by the potentials $E_0$ and
$z_0$ respectively. Thus, one can obtain an initial EMDA solution using
two Ernst ones with ${\cal{M}}_0\neq 0$ and ${\cal{D}}_0\neq 0$.

Let us now consider the action of the $U(1,1)$ subgroup on the initial set
of the charge variables with arbitrary values of ${\cal{M}}_0$ and
${\cal{D}}_0$. The operator $T_{(0)}$, accordingly to (27) and (29), shifts
the phases of the charges to the same value $2\lambda ^0$, and preserves the
set under consideration. It provides a trivial action on the charge space
(but the corresponding transformation in the target space is nontrivial, see
(20)), and will not be used later. The operator $T_{(3)}$, that belongs to
$SU(1,1)$, can be rewritten in the form
\begin{eqnarray}
\hat T_{(3)}=\left (\begin{array}{crc}
F_1&F_2\\
\bar F_2&\bar F_1
\end{array}\right ),
\end{eqnarray}
where $F_1=f_1+i\lambda ^3f_2$ and $F_2=f_2(\lambda ^1+i\lambda ^2)$. Then
the resulting from (27) and (30) formulae
\begin{eqnarray}
{\cal M} = F_1^2{\cal M}_0-\bar F_2^2{\cal D}_0,
\qquad
{\cal D} = \bar F_1^2{\cal D}_0-F_2^2{\cal M}_0,
\qquad
{\cal Q} = F_1F_2{\cal M}_0-\bar {F_1}\bar {F_2}{\cal D}_0
\end{eqnarray}
show the generation of the electromagnetic charge ${\cal Q}$.
The following from (41) relations, which express the initial charges
${\cal M}_0$ and ${\cal D}_0$ and the transformation parameters $\lambda ^i$,
combined into the functions $F_1$ and $F_2$, in terms of the resulting charges
${\cal M}$, ${\cal D}$ and ${\cal Q}$, are:
\begin{eqnarray}
[\mid F_1\mid ^2+\mid F_2\mid ^2]^{-1}\bar F_1F_2=
[{\cal M}\bar {\cal M}-{\cal D}\bar {\cal D}]^{-1}
[{\cal Q}\bar {\cal M}+\bar {\cal Q}{\cal D}],
\end{eqnarray}

\begin{eqnarray}
{\cal M}_0 = [\mid F_1\mid ^2+\mid F_2\mid ^2]^{-1}[\bar F_1^2{\cal M}
+\bar F_2^2{\cal D}],
\qquad
{\cal D}_0 = [\mid F_1\mid ^2+\mid F_2\mid ^2]^{-1}[F_2^2{\cal M}
+F_1^2{\cal D}].
\end{eqnarray}

It is convenient to introduce the new set of the $SU(1,1)$
group parameters, using the evident relation $\mid F_1\mid ^2-\mid F_2\mid ^2
=1$:
\begin{equation}
F_1=\cosh \tilde \alpha e^{i\tilde \lambda ^3},
\qquad
F_2=\sinh \tilde \alpha e^{i(\tilde \lambda ^3+\tilde \beta)}.
\end{equation}

As it follows from (41), the parameter $\tilde \lambda ^3$ provides the
equal to $2\tilde \lambda ^3$ phase shift for ${\cal M}_0$ and the phase
shift to $-2\tilde \lambda ^3$ for ${\cal D}_0$, which can be removed by
the appropriate re--definition of the initial values of these phases. Thus,
this parameter can be put equal to zero without the loss of generality for
the charge generation technique (the target space transformation
corresponding to $\tilde \lambda ^3$ coincides with the one written
in (21) with the parameter $\lambda ^3 = \tilde \lambda ^3$ in the
case of $\tilde \alpha = \tilde \beta = 0$).

Finally, the non--trivial $U(1,1)$ group action on the chosen charge space
subset is connected with two nonvanishing parameters $\tilde \alpha =
\alpha$ and $\tilde \beta = \beta$. The corresponding action of the isometry
subgroup on the target space is given by (22).

It can easily be shown using the relation (42) and the inequality
$\tanh 2\alpha < 1$, that the presented generation technique allows to
obtain the charges, which values satisfy the more stronger inequality
$\mid {\cal M}\bar {\cal M}-{\cal D}\bar {\cal D}\mid -
2\mid {\cal Q}\bar {\cal M}+\bar {\cal Q}{\cal D}\mid >0$, than the
Bogomol'nyi bound $\bar {\cal M}{\cal M}+\bar {\cal D}{\cal D}-
2\bar {\cal Q}{\cal Q} > 0$. A special case of $\mid {\cal M}\mid
= \mid {\cal D}\mid$ rises when $\mid {\cal M}_0\mid =\mid {\cal D}_0\mid$,
and corresponds to positive values of the invariants $I_1$ and $I_2$.

The performance of the described generation procedure leads to the charge
matrix $\hat Q=\hat Q_1+\hat Q_2$, where, accordingly to (24) (41) and (44),
\begin{eqnarray}
\hat Q_1 &=& T^TQ_{10}T={\cal M}_0\left (\begin{array}{ccc}
\cosh^2\alpha &\quad&\sinh\alpha \cosh\alpha e^{i\beta}\\
\sinh\alpha \cosh\alpha e^{i\beta}&\quad&\sinh^2\alpha e^{2i\beta}
\end{array}\right ),
\nonumber\\
\hat Q_2 &=& T^TQ_{20}T=-{\cal D}_0\left (\begin{array}{ccc}
\sinh^2\alpha e^{-2i\beta}&\quad&\sinh\alpha \cosh\alpha e^{-i\beta}\\
\sinh\alpha \cosh\alpha e^{-i\beta}&\quad&\cosh^2\alpha
\end{array}\right ).
\end{eqnarray}

Charge transformations for the stationary heterotic string theory with
moduli fields in the classical and quantum cases has been analized in
\cite {cy2}--\cite {sen2}.
\section{Double Rotating Dyon}
Let us take as the initial solution the double
Kerr--NUT one, i.e.,
\begin{eqnarray}
E_0 = i(1-2{\cal M}_0R_{01}^{-1}),
\qquad
z_0 = i(1-2{\cal D}_0R_{02}^{-1}),
\end{eqnarray}
where $R_{0k}=r_k+i(\nu_k+a_k\cos\theta_k)$; $k=1,2$. Here the two coordinate
sets are connected with the polar system as
\begin{eqnarray}
\rho = [(r_k-\mu_k)^2+b_k^2]^{\frac{1}{2}}\sin\theta_k,
\qquad
z = z_k+(r_k-\mu_k)\cos\theta_k,
\end{eqnarray}
where $z_k$ denotes the locations of the sources. The constants $\mu_k$ and
$\nu_k$ are, correspondingly, the real and imaginary parts of the complex
charges,
\begin{equation}
{\cal M}_0=\mu_1+i\nu_1,\quad {\cal D}_0=\mu_2+i\nu_2,
\end{equation}
and $b_k^2=\mu_k^2+\nu_k^2-a_k^2$.

The general formula (22) application to (46)--(48) leads to the
following result:
\begin{equation}
Z=i[(1-2\epsilon)\sigma_3-2\hat Q_1R_1^{-1}-2\hat Q_2R_2^{-1}].
\end{equation}
Here two modified radial functions  entered according to
\begin{eqnarray}
R_1^{-1} &=& H^{-1}R_{01}^{-1}(1-{\cal D}_0R_{02}^{-1}),
\nonumber\\
R_2^{-1} &=& H^{-1}R_{02}^{-1}(1-{\cal M}_0R_{01}^{-1}).
\end{eqnarray}
The introduced function $H$  tends to unit in a space
infinity,
\begin{equation}
H^{-1}=1+2i\sin\beta \tanh\alpha ({\cal Q}_1R_1^{-1}+{\cal Q}_2R_2^{-1});
\end{equation}
while the function $\epsilon$ vanishes as
\begin{equation}
\epsilon={\cal M}_0{\cal D}_0H^{-1}R_{01}^{-1}R_{02}^{-1}.
\end{equation}
Finally, the sum of the written in (45) matrices $\hat Q_1$ and $\hat Q_2$
defines the set of charges for the constructed double rotating dyon.

This solution transforms into the one describing the single rotating dyon
with $I_1 > 0$ and $I_2 = 0$ \cite {gk3}, after the removal of any one of
the parameters ${\cal M}_0$ or ${\cal D}_0$.

The solution (49)--(52) generalizes also the one obtained earlier
\cite {gk2}, which was deprived of magnetic, axion and NUT charges. Its
following generalization will be connected with the completely independent
values of the matrices $\hat Q_1$ and $\hat Q_2$, and can not be achieved
using only the isometry group transformations.

The extremal IWP--like EMDA solutions of this type were found in \cite {bko},
while the solutions describing the single rotating dyon for the low energy
heterotic string theory with moduli fields were obtained in
\cite {sen3}--\cite {cy3}.

\section{Conclusion}
The action of the isometry subgroup, which preserves the trivial values of
the fields, on the Ernst--like potentials as well as on the physical charges,
is studied for the stationary D=4 EMDA theory. The general charge generation
technique covariant on this subgroup is analyzed, and the corresponding
solution generation procedure is formulated. A solution describing the
double rotating dyon with arbitrary values of the charges is constructed.

\acknowledgments
We would like to thank our colleagues from the Nuclear Physics Institute for
an encouraging relation to our work.



\begin{references}
\bibitem{ms}
N. Marcus and J.H. Schwarz, Nucl. Phys. {\bf B228} (1983) 145.
\bibitem{dl}
M. Duff and J. Lu, Nucl. Phys. {\bf B347} (1990) 394.
\bibitem{mahs}
J. Maharana and J.H. Schwarz, Nucl. Phys. {\bf B390} (1993) 3.
\bibitem{bkr}
A.K. Biswas, A. Kumar and K. Ray, Nucl. Phys. {\bf B453} (1995) 181.
\bibitem{sen1}
A. Sen, Nucl. Phys. {\bf B447} (1995) 62.
\bibitem{mah}
J. Maharana, IP--BBSR--95--5, hep--th/9502002.
\bibitem{ggk1}
D.V. Gal'tsov, A.A. Garcia and O.V. Kechkin, J. Math. Phys. {\bf 36} (1995)
5023.
\bibitem{gk1}
D.V. Gal'tsov and O.V. Kechkin, Phys. Rev. {\bf D54} (1996) 1656.
\bibitem{ky1}
O. Kechkin and M. Yurova, hep--th/9604175.
\bibitem{gk2}
D.V. Gal'tsov and O.V. Kechkin, Phys. Lett. {\bf B361} (1995) 52.
\bibitem{gk3}
D.V. Gal'tsov and O.V. Kechkin, Phys. Rev. {\bf D50} (1994) 7394.
\bibitem{ggk2}
A. Garcia, D. Galtsov and O. Kechkin, Phys. Rev. Lett. {\bf 74} (1995) 1276.
\bibitem{iw}
W. Israel and G.A. Wilson, J. Math. Phys. {\bf 13} (1972) 865.
\bibitem{cy1}
M. Cveti$\check c$ and D. Youm, Phys. Rev. {\bf D53} (1996) 584.
\bibitem{stw}
A. Shapere, S. Trivedi and F. Wilczek, Mod. Phys. Lett. {\bf A6} (1991) 2677.
\bibitem{cy2}
M. Cveti$\check c$ and D. Youm, IASSNS--HEP--95/107, PUPT--1582 hep-th/9512127.
\bibitem{sen2}
A. Sen, Nucl. Phys. {\bf B434} (1995) 179.
\bibitem{bko}
E. Bergshoeff, R. Kallosh and T. Ortin, UG--3/96, SU--ITP--19,
CERN--TH/96--106, hep--th/9605059.
\bibitem{sen3}
A. Sen, Nucl. Phys. {\bf B440} (1995) 421.
\bibitem{hs}
G. P. Horowitz and A. Sen, Phys. Rev. {\bf D53} (1996) 808.
\bibitem{cy3}
M. Cveti$\check c$ and D. Youm, IASSNS--HEP--96/43, UPR--700--T, PUPT--1623
\\hep--th/9605051.
\end{references}
\end{document}